\documentclass[aps,prl,twocolumn,superscriptaddress,shopacs]{revtex4}
\usepackage{graphicx}
\usepackage{bm}
\usepackage{epstopdf}
\usepackage{multirow}
\usepackage{float}
\usepackage{tabularx}
\usepackage{color}
\usepackage{xr}
\usepackage{amsmath}
\usepackage{color}
\usepackage{ulem}

\begin{document}

\title{Weak phonon coupling to nematic quantum critical mode in BaFe$_2$(As$_{1-x}$P$_x$)$_2$}

\author{S. Wu}
\email{swu13@scu.edu}
\affiliation{Department of Physics, University of California, Berkeley, California, 94720, USA}
\affiliation{Material Sciences Division, Lawrence Berkeley National Lab, Berkeley, California, 94720, USA}
\affiliation{Department of Physics and Engineering Physics, Santa Clara University, Santa Clara, CA, 95053}

\author{D. Ishikawa}
\affiliation{Materials Dynamics Laboratory, RIKEN SPring-8 Center, Sayo, Hyogo 679-5148, Japan}
\affiliation{Precision Spectroscopy Division, SPring-8/JASRI, 1-1-1 Kouto, Sayo, Hyogo, 679-5198, Japan}

\author{A. Q. R. Baron}
\affiliation{Materials Dynamics Laboratory, RIKEN SPring-8 Center, Sayo, Hyogo 679-5148, Japan}
\affiliation{Precision Spectroscopy Division, SPring-8/JASRI, 1-1-1 Kouto, Sayo, Hyogo, 679-5198, Japan}

\author{A. Alatas}
\affiliation{Advanced Photon Source, Argonne National Laboratory, Argonne, Illinois 60439, USA}

\author{A. H. Said}
\affiliation{Advanced Photon Source, Argonne National Laboratory, Argonne, Illinois 60439, USA}

\author{Jiayu Guo}
\affiliation{Center for Correlated Matter and School of Physics, Zhejiang University, Hangzhou 310058, China}

\author{Y. He}
\affiliation{Department of Applied Physics, Yale University, New Haven, Connecticut 06511, USA}

\author{X. Chen}
\affiliation{Department of Physics, University of California, Berkeley, California, 94720, USA}
\affiliation{Center for Neutron Science and Technology, School of Physics, Sun Yat-Sen University, Guangzhou, Guangdong, 510275, China}

\author{Y. Song}
\affiliation{Center for Correlated Matter and School of Physics, Zhejiang University, Hangzhou 310058, China}

\author{J. G. Analytis}
\affiliation{Department of Physics, University of California Berkeley, California, 94720, USA}
\affiliation{CIFAR Quantum Materials, CIFAR, Toronto, Ontario M5G 1M1, Canada}

\author{Dung-Hai Lee}
\affiliation{Department of Physics, University of California, Berkeley, California, 94720, USA}
\affiliation{Material Sciences Division, Lawrence Berkeley National Lab, Berkeley, California, 94720, USA}

\author{R. J. Birgeneau}
\email{robertjb@berkeley.edu}
\affiliation{Department of Physics, University of California, Berkeley, California, 94720, USA}
\affiliation{Material Sciences Division, Lawrence Berkeley National Lab, Berkeley, California, 94720, USA}

\date{\today}
\begin{abstract}
In this work, we investigate the softening of the in-plane transverse acoustic phonon driven by electronic nematicity in BaFe$_2$(As$_{1-x}$P$_x$)$_2$ using inelastic X-ray scattering, with a focus on the optimally doped sample ($x = 0.31$) sample — a system exhibiting signatures of a putative nematic quantum critical point and minimal disorder among iron pnictides. We observe only a modest softening of the phonon frequency and no evidence of critical damping, suggesting that the nematic quantum critical fluctuations couple only weakly to the lattice from our quantum critical model. Given the close proximity of the structural and magnetic transition temperatures in the underdoped sample — which implies that spin-nematic fluctuations couple strongly to the lattice — we conjecture that the quantum critical nematic fluctuations are predominantly orbital in origin.
\end{abstract}

\maketitle

Quantum criticality is an essential feature of a wide range of exotic electronic behaviors.
Iron-based superconductors presents a prominent example; in particular, fluctuations near the nematic quantum critical point (QCP) are closely related to unconventional superconductivity and anomalous electronic behaviors \cite{lederer2015, metlitski2015,samuel2017,labat2017, Metzner2003,fernandes2014,fradkin2010,Chu2012,hosoi2016,Hayes2016,Worasaran2021-lw,osti_1779222,Palmstrom2022,nakai2010,shibauchi2014,Analytis2014,Kuo2016,hashimoto2012}. Among them, BaFe$_2$(As$_{1-x}$P$_x$)$_2$ (BFAP) has emerged as a model system for exploring the complex interplay between intertwined electronic phases, displaying a rich tapestry of phenomena that continue to fuel research interest; these include electronic and structural nematicity, antiferromagnetic order, high-temperature superconductivity, Non-Fermi Liquid behavior, and a putative nematic QCP near the optimal doping ratio \cite{kasahara2010,shishido2010,hashimoto2012,walmsley2013,Analytis2014,Kuo2016}.
In this work, we focus on the study of the coupling between lattice dynamics and electronic nematicity, with particular attention to its behavior near the putative QCP.

A central question concerns the leading microscopic origin of electronic nematicity. Two principal mechanisms have been proposed \cite{fernandes2014np,fernandes2012,stanev2013,kang2011}: the orbital-driven scenario, in which nematicity arises from unequal occupation of the Fe $3d_{xz}$ and $3d_{yz}$ orbitals, and the spin-driven scenario, in which the nematic order emerges from the system’s spontaneous selection of an antiferromagnetic wavevector - $(\pi,0)$ or $(0,\pi)$ - thus breaking the fourfold rotational symmetry of the lattice. The dominant mechanism remains unsettled and appears to vary across different families of iron-based superconductors \cite{chubukov2016}.

Evidence for quantum critical phenomena in iron-based superconductors has been limited thus far. BFAP stands out as a compelling case, exhibiting strong signatures of an electronic nematic QCP near the critical doping level $x\approx 0.31$ \cite{Kuo2016,Analytis2014, hashimoto2012, kasahara2010, shishido2010, walmsley2013,shibauchi2014}. Unlike its carrier-doped counterparts, BFAP undergoes isovalent substitution, resulting in significantly reduced disorder and impurity scattering \cite{Abrahams_2011}. This positions BFAP as a rare and clean platform in which nematic fluctuations can be continuously tuned through criticality.

The interplay between electronic nematicity and magnetism is salient in the underdoped regime of BFAP, where the system exhibits a N\'eel-type spin-density wave order accompanied by a tetragonal-to-orthorhombic structural phase transition \cite{Jiang_2009,kasahara2010}. This structural nematicity — manifested as a lattice distortion that breaks the crystal’s fourfold rotational symmetry — is widely interpreted to be spin-driven nematicity in both BFAP and electron-doped materials \cite{chubukov2016}, as evidenced by the close proximity of the structural ($T_s$) and magnetic transition temperatures ($T_N$) in both compounds \cite{iye2012,kasahara2010,Hardy_2010,kim2011}.

\begin{figure*}
\includegraphics[width=2\columnwidth,clip,angle =0]{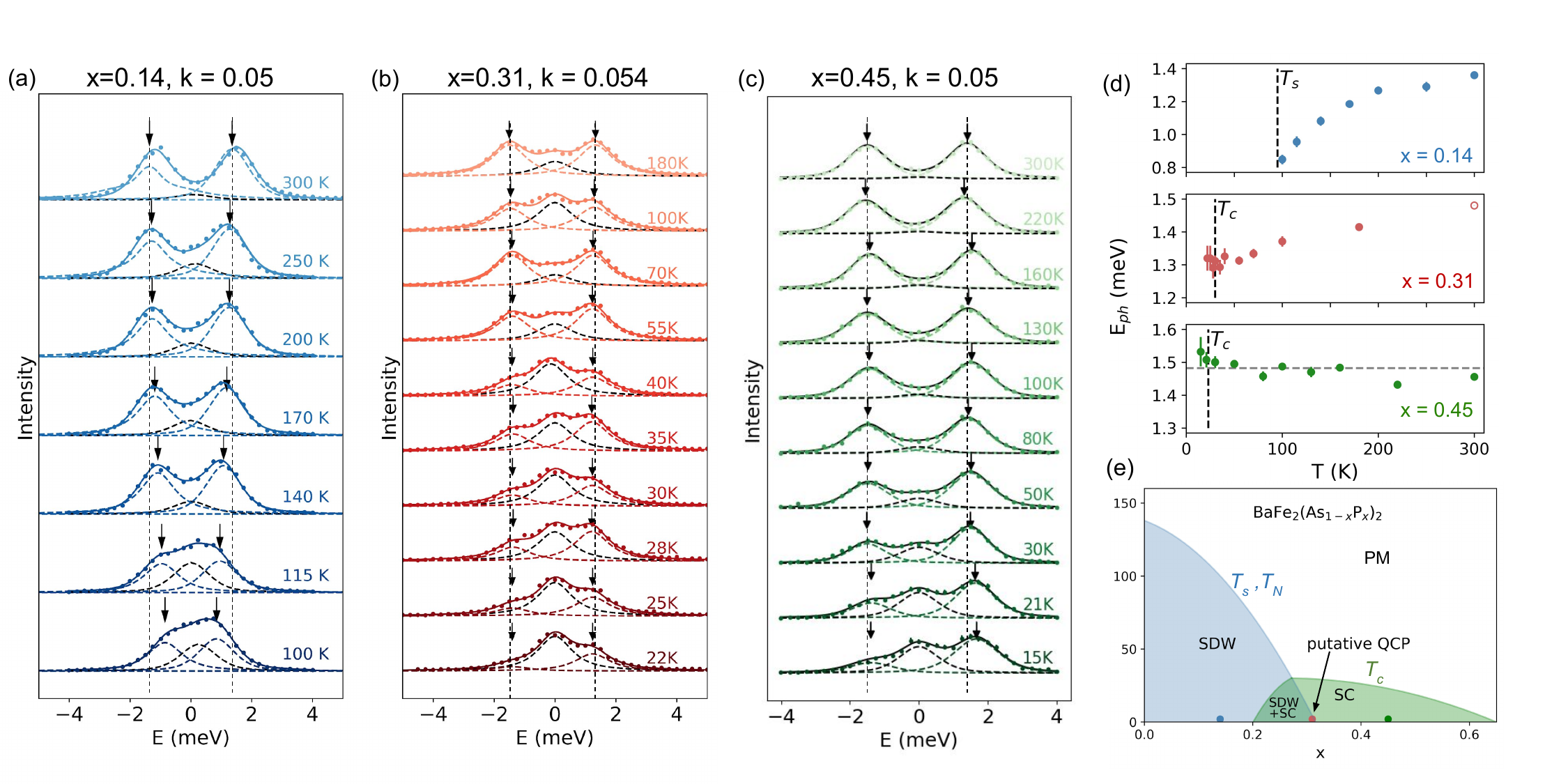}
\caption{ (a-c) Temperature-dependent energy scans at wave vector ${\bf Q}=(4,k,0)$ for BaFe$_2$(As$_{1-x}$P$_x$)$_2$ (BFAP) with x = 0.14 and $k$ = 0.05, $x$ = 0.31 and $k$ = 0.054, x = 0.45 and $k$ = 0.05. Solid symbols and dashed curves are experimental data and the corresponding fits, respectively. Black arrows mark the extracted phonon energies $E_{\rm ph}$ for the Stokes and anti-Stokes phonons. The vertical dashed lines are aligned to the phonon energies at the highest measured temperature. The data for x = 0.14 and 0.45 are from APS, $x$ = 0.31 are from SPring-8. Temperature dependence of the fit values of $E_{\rm ph}$ in panels (a-c) are respectively shown in (d).  The empty dot for $x$ = 0.31 sample at room temperature is obtained by extrapolation of phonon dispersion from other $k$s (Fig. 3(a)). The horizontal grey dashed line for the plot in the x = 0.45 sample denotes the averaged phonon energy. Vertical error bars are least-square fit errors of 1 s.d. (e) Schematic phase diagram of BaFe$_2$(As$_{1-x}$P$_x$)$_2$, showing the orthorhombic spin density wave (SDW-O), tetragonal paramagnetic (PM-T), and superconducting (SC) tetragonal phases (adapted from Ref. \cite{hashimoto2012}). Blue, red and green dots represent the three samples studied in this work. $x$ = 0.31 sample locates around the putative nematic quantum critical point region.  }
\label{phonon}
\end{figure*}

More importantly, in BFAP both magnetic order and structural distortion disappear in a weakly first-order manner \cite{iye2012,hu2015,allred2014,bohmer2012}, suggesting an avoided magnetic QCP similar to that observed in its electron-doped counterparts \cite{lu2013}. In contrast, electronic nematic fluctuations, inferred from elastoresistivity, persist beyond the critical ratio $x\approx 0.31$ and decay only gradually with further doping \cite{Chu2012,Kuo2016}. This disparity raises key questions about the underlying leading driving mechanism of electronic nematicity in the quantum critical regime.

Phonon energy and linewidth serve as sensitive probes of the nematoelastic coupling between the lattice and electronic nematicity \cite{niedziela2011,parshall2015,liyu2018,weber2018,weber2020,merritt2020,wu2021}. If the nematic QCP in BFAP is spin-driven, as suggested by the strong coupling between structural distortion and spin fluctuations in the underdoped phase, enhanced nematoelastic coupling can lead to pronounced softening of the acoustic phonon energy and broadening of the linewidth due to increased nematic susceptibility and scattering with critical electronic modes. Thus, probing the lattice dynamics, specifically the in-plane transverse acoustic phonon via high-resolution inelastic X-ray scattering provides a powerful approach to uncover the microscopic nature of the nematic QCP in BFAP, a model system for a clean, isovalently substituted iron-based superconductor hosting a nematic quantum criticality.

In this paper, we report inelastic X-ray scattering measurements in BaFe$_2$(As$_{1-x}$P$_x$)$_2$ at three isovalent phosphorus substitutions $x$: an underdoped sample (UD, $x = 0.14$) exhibiting a paramagnetic tetragonal to antiferromagnetic orthorhombic phase at T$\rm _s$ = 95 K, an optimally-doped sample (OP, $x_c = 0.31$) with superconducting temperature T$\rm _c$ = 30 K, and an over-doped sample (OD, $x = 0.45$)  with T$\rm _c$ = 23 K  (Fig. \ref{phonon} (e)). At the critical doping  $x = 0.31$,  we observe weak phonon softening — with no evidence of phonon damping by the quantum critical modes — and vanishing phonon softening for $x = 0.45$. These results point to a remarkably weak coupling between quantum nematic fluctuations and the lattice in $x = 0.31$  sample from our model, sharply contrasting with expectations from the spin-driven QCP scenario. Our work indicates a predominantly orbital-driven origin of the nematic quantum criticality in this system, only weakly coupled to the magnetism and structural distortion.

\begin{figure}
\includegraphics[width=1.\columnwidth,clip,angle =0]{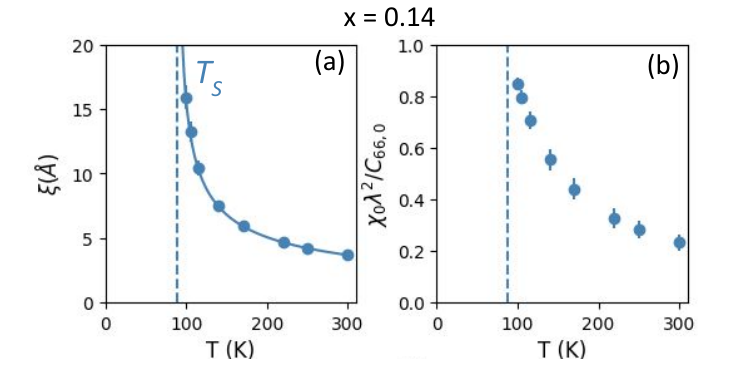}
\caption{\label{UD} (a-b) Nematic correlation length $\xi$ and derived bare nematic susceptibility in units of $\lambda^2/C_{66,0}$ as a function of temperature for BaFe$_2$(As$_{1-x}$P$_x$)$_2$ with $x=$ 0.14. The solid lines are fits by using Curie-Weiss mean field model as described in previous work \cite{wu2021} and the supplementary information.  All vertical error bars are least-square fit errors of 1 s.d.}
\label{UD}
\end{figure}
\begin{figure*}
\includegraphics[width=2.\columnwidth,clip,angle =0]{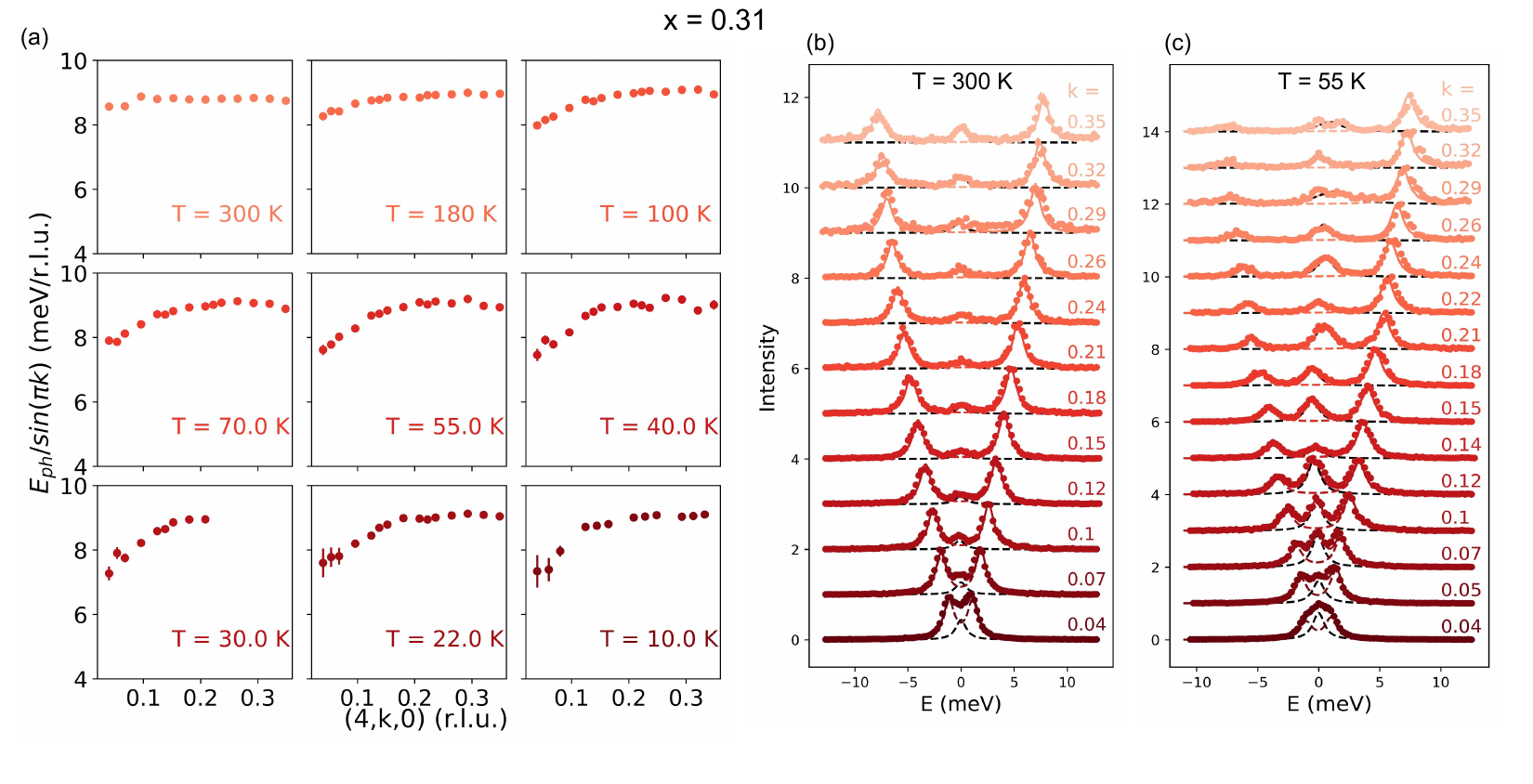}
\caption{\label{OP} (a) Extracted $E_{\rm ph}/k$ plotted against $k$ (solid symbols) for BaFe$_2$(As$_{1-x}$P$_x$)$_2$ with $x=$ 0.31 from the energy scans at $\bf {Q}$ $= (4,k,0)$ by using quantitative fitting methods from previous studies \cite{wu2021}.
(b-c) Wave-vector dependent energy scans and corresponding fits at $T$ = 300 K and  55 K. The solid lines are fits by using Gaussian quantum critical model described in the text.  
All vertical error bars are either least-square fit errors of 1 s.d., or obtained through the propagation of fit errors. 
}
\label{OP}
\end{figure*}

High-quality single crystals were synthesized using a self-flux method as previously described \cite{analytis2010,Hayes2016}. The size of the crystals are small, far below the size needed for  neutron scattering. Therefore, phonon measurements were performed using high-resolution inelastic X-ray scattering techniques carried out at the RIKEN Quantum NanoDynamics beamline, BL43XU, of SPring-8 \cite{ishikawa2015,baron2020introduction,baron2010,Ishikawa2021}, Japan for the critical doping sample $x$ = 0.31, and at the 30-ID beamline \cite{Toellner,Said} at the Advanced Photon Source (APS), Argonne National Laboratory for the other two samples.  The incident photon energies were fixed at 21.75 keV at BL43LXU and 23.7 keV at the 30-ID beamline, respectively.  The energy resolution was analyzer dependent, varying from 1.25 meV to 1.6 meV FWHM for the analyzers used in the experiment. Momentum transfer ${\bf Q}$ is referenced in reduced lattice units, using the tetragonal 2-Fe unit cell with in-plane lattice constants of  $3.93$~${\rm \AA}$ and $3.92$~${\rm \AA}$. More experimental details are referred to the supplementary information.

In our measurements of the in-plane transverse acoustic phonons at the wave vector transfer $\mathbf{Q}$ = $(4,k,0)$, we observed discernible phonon softening upon cooling for under-doped ($x$ = 0.14) and moderate softening in critical doped ($x$ = 0.31) samples, as shown in Figs. \ref{phonon} (a) and (b). This trend is further supported by the extracted phonon energies, $E_{\rm ph}$, obtained from the Stokes and anti-Stokes \cite{weber2018,weber2020,wu2021} phonons as a function of temperature at fixed $k$s, presented in Fig. \ref{phonon} (d). The fitting methods used for this analysis are described in \cite{wu2021}.

In the $x$ = 0.14 sample, $E_{\rm ph}$ decreases gradually from 1.4 meV at $\rm T = 300$ K to 0.85 meV as the temperature approaches $T_{\rm S} \sim$  95 K at $k$ = 0.05, reflecting a substantial 40\% reduction in phonon energy. 
In comparison, the $x$ = 0.31 sample shows a more moderate decrease from 1.5 meV at $\rm T = 300$ K to 1.3 meV below 55 K, and toward $T_{\rm c}$ = 30 K at $k$ = 0.054. This 13\% phonon softening is much less pronounced than in the $x$ = 0.14 sample. 
The reduced softening in the $x=$ 0.31 sample is also evident in the $k$-dependent sound velocity plots (Fig. S1 \&  Fig. 3 (a)), which show a smaller value 
at the lowest measured wave-vector. 
In contrast, the over-doped $x$ = 0.45 sample displays negligible phonon softening down to $T = 15$ K ($T_c = 23$ K), as shown in Figs. \ref{phonon}(c) and (e). The phonon energy remains at 1.5 meV at $k$ = 0.05, indicating a lack of appreciable electron-phonon coupling in this regime. While a slight hardening below $T_{\rm c}$ cannot be entirely ruled out, the reduced anti-Stokes intensity at low temperatures makes a definitive conclusion difficult. 

The $x = 0.14$ sample undergoes a tetragonal-to-orthorhombic structural transition at $T_s$. To assess the momentum-dependent phonon softening, we analyze the $k$-dependent sound velocity, $E_{\rm ph}/k$, as a function of temperature  using the exact mean-field approach described in Ref. \cite{wu2021}. A global fit of the data to this thermal nematic model was performed across the full temperature range, with the resulting fits shown in Fig. S1. The extracted temperature dependence of the nematic correlation length $\xi$ and the bare nematic susceptibility $\chi_0$ (expressed in units of $\lambda^2 / C_{66,0}$) are shown in Fig. \ref{UD}. The enhancement of $\chi_0$ approaching $T_s$ agrees with the interpretation of nematic-driven structural distortion, likely originating from spin nematicity.

Encouraged by the success of mean-field theory in capturing the thermal nematic criticality, and the fact that the nematic fluctuation at QCP is described by quantum Ginzburg-Landau theory in 3+1 (the upper critical) dimension, we adopt a similar Gaussian framework to describe the quantum critical regime. This approach is further justified by the fact that coupling to acoustic phonons mediates long-range interactions.
We consider the critical nematic field ($\phi$) interacting with the acoustic phonon mode ($u$) via the nemato-elastic coupling strength $\lambda$. The corresponding imaginary-time Lagrangian can be expressed as:
\begin{equation}
L = (\partial_\tau \phi)^2 +\sum_{j=1}^3 (\partial_j \phi)^2 + m^2 \phi^2 + \lambda \phi u + (\partial_\tau u)^2 + v^2\sum_{j=1}^3(\partial_j u)^2,
\end{equation}
where $v$ represents the sound velocity. In Eq. (1), we have appropriately scaled the spatial and temporal coordinates, as well as the field $\phi$, so that the coefficients of corresponding terms are unity. Similarly, $u$ has been scaled such that the coefficient  of  $(\partial_\tau u)^2$ is also unity. The parameter in Eq. (1) is proportional to the inverse nematic correlation length $\xi$. 


Integrating out the nematic field $\phi$ and Wick rotate to real frequency leads to a the following equation determining the renormalized phonon dispersion relation. 
\begin{equation}
-\omega^2 + v^2 q^2 - \frac{\lambda^2}{-\omega^2 + q^2 + m^2} = 0.
\label{rf}
\end{equation}
Here $q$ is the wave vector. The 3rd term in Eq. (2) originates from the nematic fluctuations with the velocity of the nematic wave equal to unity as discussed above.  A prediction of  Eq.(\ref{rf}) is that both the phonon frequency and line-width exhibit the following scaling behavior:
\begin{align}
\omega_q  = \frac{1}{\xi} F_R(q\xi, \lambda \xi^2) \\
\Gamma_q  = \frac{1}{\xi} F_I(q\xi, \lambda \xi^2),
\end{align}
where $F_R$ and $F_I$ are scaling functions. 
One implication of Eq. (3) and (4) is that after all the scaling performed on Eq. (1), both the real and imaginary part of the phonon frequency have unit $1/\xi$ and $\lambda$ has unit $1/\xi^2$.
Solving Eq.(\ref{rf}) yields the renormalized phonon frequency $\omega_q$, where $\omega_q = \omega_{ph} + i\Gamma$. Here, the real part corresponds to the phonon energy ($\omega_{ph}$), and the imaginary part ($\Gamma$) represents phonon damping width, which can both determined by a single parameter $\lambda$ in Eq. \ref{rf}.  Consequently, the phonon structure factor takes the form of a damped harmonic oscillator:
\begin{equation}
S(q,\omega) = (n_{\omega} + 1)L(\omega - \omega_{ph}, \Gamma) - n_{\omega}L(\omega + \omega_{ph}, \Gamma)
\end{equation}
where $n_{\omega}$ is the Bose factor, and $L(\omega \pm \omega_{ph}, \Gamma)$ denotes the Lorentzian functions representing the Stokes and anti-Stokes phonon. The resulting intensity is subsequently convolved with the instrumental resolution function $R_s(\omega)$.

We determine the bare phonon velocity $v$ by fitting Eq.(2) and (5) with $\lambda=0$ to our highest temperature spectra at $T =$ 300 K. The fits also included the phonon intensities and central elastic peak intensities. The resulting fits are shown in Fig. \ref{OP}(b) for varying $k$ values, yielding $v = 8.8(1)$ meV/r.l.u. These fits agree well with the experimental data, with a reduced $\chi^2$ value of 2.5, although there is some deviation closer to the zone boundary due to the lattice dynamics and anharmonicities.

To highlight the electron-phonon coupling, we fit our lower temperature (down to 55 K) data with fixed $v$ (determined from the $T =$ 300 K data)  and finite $\lambda$ as shown in Fig. \ref{OP}(c), with a reduced $\chi^2$ value of 3 and a fitted coupling strength parameter of $\lambda = 3.2(1)$. 
Compared to the 300 K case, introducing a finite $\lambda$ leads to moderate phonon softening, but without the phonon frequency approaching zero or exhibiting phonon damping. 
As the temperature decreases, the phonon dispersion spectra remain qualitatively comparable, with no additional phonon softening observed or broadening of the phonon width (Fig. 1(d)). However, at lower temperatures, suppressed phonon intensities lead to poorer fits at larger $k$ values when applying a linear dispersion model. For this reason, those fits are not shown.

To better illustrate the impact of the coupling strength, we calculated the acoustic phonon intensities at $(4, k, 0)$, along with the phonon frequency $\omega_{ph}$ and line-width $\Gamma$ for various values of $\lambda$ (Fig. S2). Increasing $\lambda$ from 3.2 to 5 leads to greater phonon softening without damping ($\Gamma = 0$) at low temperatures. At $\lambda = 10$, the mode softens further toward zero frequency with significant broadening at low temperatures, reflecting diverging nematic correlation length and fluctuation time near the nematic QCP. These results demonstrate that the manifestation of QCP through phonon softening and damping depends on the coupling strength between nematic fluctuations and the lattice.

We observe the absence of a vanishing phonon frequency and negligible phonon damping at low temperatures in the $x=$ 0.31 sample near the nematic quantum critical point. 
Both our fits and theoretical calculations suggest a weak coupling between the nematic critical fluctuations and the acoustic phonon modes. 
In the $x$ = 0.14 sample, where nematicity is likely spin-driven, strong phonon softening is observed as the system approaches a thermally driven structural phase transition. In contrast, the $x$ = 0.31 sample exhibits considerably weaker phonon softening, despite the presence of quantum critical nematic fluctuations revealed through elastoresistivity measurements \cite{Chu2012, Kuo2016}. This disparity suggests that quantum critical nematic fluctuations are primarily mediated via the orbital channel, and orbital nematic fluctuations couple weakly to the lattice. Upon further P substitution into the overdoped regime, the phonon softening becomes negligible across a wide temperature range, while elastoresistivity continues to indicate moderate electronic nematic fluctuations. This suggests that the nematic fluctuations in this doping range are also orbital-driven.

It is important to note that the strength of coupling between orbital-driven nematicity and the lattice may not be universal across different families of iron-based superconductors. For example, in FeSe$_{1-x}$S$_x$\cite{hosoi2016}, the structural distortion occurs in the absence of magnetic order, suggesting that the coupling between orbital nematicity and the lattice can be significantly stronger in that system. Additionally, our results do not directly address whether the non-Fermi-liquid transport behavior observed in this material arises from critical nematic fluctuations. Both spin-driven and orbital-driven nematic fluctuations can, in principle, contribute to non-Fermi-liquid transport. Therefore, the absence of phonon softening beyond $x$ = 0.31 does not exclude the presence of orbital-driven nematic fluctuations; rather, it indicates an approximate decoupling from spin-driven nematicity. Finally, we note that in the $x=$ 0.31 sample, which exhibits the highest $T_c$, the phonon softening halts below the superconducting transition temperature. A more comprehensive understanding of the interplay between superconductivity and nematic criticality remains an open question.

In summary, we investigated the coupling between electronic nematic fluctuations in isovalently doped BaFe$_2$(As$_{1-x}$P$_x$)$_2$ using inelastic X-ray scattering over a range of phosphorus concentrations, targeting a system that exhibits signatures of a nematic quantum critical point. In the underdoped sample, we observed softening of the in-plane acoustic phonon mode associated with the tetragonal-to-orthorhombic structural phase transition, and extracted the nematic susceptibility and correlation length using an exact mean-field analysis as in prior work. In the optimally doped sample ($x_c = 0.31$) near the quantum critical doping of nematicity, we did not observe either critical phonon softening or anomalous damping. Using a Gaussian quantum critical model incorporating nematoelastic coupling, we attribute these to the weak coupling between nematic fluctuations and the lattice. This interpretation is further supported by the disappearance of phonon softening in the overdoped regime, suggesting even weaker coupling between the nematic fluctuations observed in elasoresistivity measurement and the lattice. These lead to our conjecture: the nematic fluctuations in the $x$ = 0.31 and overdoped samples are likely driven more by orbital rather than spin.

 
This work was  funded by the U.S. Department of Energy, Office of Science, Office of Basic Energy Sciences, Materials Sciences and Engineering Division under Contract No. DE-AC02-05-CH11231 within the Quantum Materials Program (KC2202). Part of work was also supported by Santa Clara University grant. The work at Zhejiang University is supported by the National Key R\&D Program of China (Grant No. 2024YFA1409200). This research used resources of the Advanced Photon Source, a U.S. Department of Energy (DOE) Office of Science User Facility operated for the DOE Office of Science by Argonne National Laboratory under Contract No. DE-AC02-06CH11357. The data collected at the RIKEN Quantum NanoDynamics beamline, BL43XU of SPring-8, is available upon request under proposal number 2021A1094 (application no: 49864).

\bibliography{bibfile_new}
\end{document}